\documentclass[runningheads]{llncs}
\usepackage[T1]{fontenc}
\usepackage{graphicx}
\usepackage{multirow}
\begin{document}
%
\title{Segmentation of Aortic Vessel Tree in CT Scans with Deep Fully Convolutional Networks}
\titlerunning{Segmentation of Aortic Vessel Tree in CT Scans}
%
\author{Shaofeng Yuan\inst{1} \and Feng Yang\inst{2}}
\authorrunning{S. Yuan and F. Yang}
%
\institute{Institute of Artificial Intelligence, Insight Lifetech, Shenzhen, China
\email{shaofeng.yuan.smu@gmail.com} \and
School of Biomedical Engineering, Southern Medical University, Guangzhou, China
}
\maketitle              
%
\begin{abstract}
Automatic and accurate segmentation of aortic vessel tree (AVT) in computed tomography (CT) scans is crucial for early detection, diagnosis and prognosis of aortic diseases, such as aneurysms, dissections and stenosis. However, this task remains challenges, due to the complexity of aortic vessel tree and amount of CT angiography data. In this technical report, we use two-stage fully convolutional networks (FCNs) to automatically segment AVT in CTA scans from multiple centers. Specifically, we firstly adopt a 3D FCN with U-shape network architecture to segment AVT in order to produce topology attention and accelerate medical image analysis pipeline. And then another one 3D FCN is trained to segment branches of AVT along the pseudo-centerline of AVT. In the 2023 MICCAI Segmentation of the Aorta (SEG.A.) Challenge , the reported method was evaluated on the public dataset of 56 cases. The resulting Dice Similarity Coefficient (DSC) is 0.920, Jaccard Similarity Coefficient (JSC) is 0.861, Recall is 0.922, and Precision is 0.926 on a 5-fold random split of training and validation set.

\end{abstract}
\section{Introduction}
The aorta is the main artery in the human body. It consists of four main parts – ascending aorta, aortic arch, descending thoracic aorta, and abdominal aorta. From the heart to the whole body, the aorta delivers oxygen-rich blood. Aortic diseases, such as aneurysms, dissections and stenoses, pose a threat to patients' life. Non-invasive examinations of the aorta are crucial for early diagnosis and timely treatment of these diseases. Segmentation of the whole aorta and all its branches, also named aortic vessel tree (AVT) segmentation, is critically important during clinical examination of non-invasive vascular imaging. However, manual segmentation of AVT is a time-consuming and labor-intensive task, and may lack reproducibility. For this task, AI-assisted automatic segmentation methods have shown to be a possible solution, which can potentially be used in clinical routine and medical research~\cite{jin2021}.

Jin et al.~\cite{jin2021} review various aorta segmentation approaches, and roughly dividing them into four classes, namely, deformable models, tracking models, deep learning models and other models. In this report, we only review the latest work based on deep learning techniques~\cite{wu2018,yuan2019,yuan2022,yuan2023}. It is well known that convolutional networks (ConvNets) and fully convolutional networks (FCNs) were greatly used in areal/volumetric medical image segmentation~\cite{ronneberger2015,milletari2016}, such as 2D/3D U-Net and 2D/3D V-Net~\cite{tang2019,tang2020}. For segmentation of AVT or partial aorta, many ConvNets/FCNs-based methods were developed~\cite{jin2021}. However, the open accessible large-scale computed tomography angiography (CTA) or magnetic resonance angiography (MRA) data with well labeled aorta are scarce, and it hinders the development of automatic AVT segmentation algorithms base on artificial intelligence (AI). The organizers of the 2023 MICCAI Segmentation of the Aorta (SEG.A.) Challenge collected 56 CTA scans from multiple centers, i.e., the KiTS Grand Challenge, the Rider Lung CT dataset, and the Dongyang Hospital~\cite{radl2022}. These scans have different pathologies, such as aortic dissections (AD) or abdominal aortic aneurysms (AAA).

For aorta segmentation, Trullo et al.~\cite{trullo2017} use the SharpMask and CRFasRNN architectures for the joint segmentation of organs at risk in CT images of the thorax, specifically the heart, esophagus, trachea and the aorta. Noothout et al.~\cite{noothout2018} propose an automatic method to segment the ascending aorta, the aortic arch and the thoracic descending aorta in low-dose chest CT using dilated convolutional networks and three orthogonal slices. Bonechi et al.~\cite{bonechi2021} use the same method but with different network architectures and pre-trained encoders for aortic segmentation in 154 CTA scans. Lin et al.~\cite{lin2022} propose a geometry-constrained deformable attention network for segmenting the aorta, including normal aorta, coarctated aorta and dissected aorta. This method is better than ~\cite{cao2019,cheng2020,chen2021,lyu2021,yu2021}.

For dissected aorta segmentation, Li et al.~\cite{li2018} formulate lumen segmentation of AD as contour extraction of aortic adventitia and intima in 2018. Cascaded ConvNets are used for contour extraction on 2D cross-section images, and 3D adventitia and intima shape models are constructed. In 2019, Cao et al.~\cite{cao2019} develop and evaluate three FCN-based models for type B AD segmentation. The serial multi-task model performed the best, in which the first network was used for the whole aorta segmentation and provides a tight bounding box for the segmentation of true or false lumen in the next network. Fantazzini et al.~\cite{fantazzini2020} extend the multi-view ConvNets~\cite{noothout2018} for segmenting the whole aortic lumen from thoracic aorta to the common iliac arteries in CTA scans. In 2020, Hahn et al.~\cite{hahn2020} develop an automated segmentation pipeline for segmentation of aortic dissection CT angiograms into TL and FL on multiplanar reformations (MPRs) perpendicular to the aortic centerline. Then they obtain aortic diameter, TL or FL cross-sectional area, and other morphologic parameters for surveillance and risk stratification. Cheng et al.~\cite{cheng2020} construct a U-Net based semantic segmentation architecture and apply it to contrast-enhanced CT images to segment the aortic true lumen in order to rapid detect AD. Chen et al.~\cite{chen2021} present a multi-stage segmentation framework for type B AD to extract true lumen (TL), false lumen (FL) and all branches. Lyu et al. ~\cite{lyu2021} propose a deep-learning-based algorithm to segment dissected aorta on CTA images combining 3D ConvNets and 2D ConvNets. Yu et al.~\cite{yu2021} implement a deep FCN framework for entire aorta segmentation, combining the U-Net model with the DenseNet model. For TL and FL identification, intersection over union tracing method is used to recognize the membrane.  In 2021, Fand et al.~\cite{fang2021} propose a fully automatic vessel analysis pipeline for dissected aorta, which can output centerlines, TL, and FL of aortic trunk and major branches. This method relies on centerline extraction. In 2022, Sieren et al.~\cite{sieren2022} use a 3D U-Net model to validate an algorithm based on deep learning for segmenting and quantifying the physiological and diseased aorta in 191 CTA exams. Xiang et al.~\cite{xiang2023} propose a flap-attention-based AD segmentation method, termed ADSeg, to automatically segment the TL, FL, and branch vessels of AD in an end-to-end manner. The ADSeg outperforms all existing methods by a large margin in terms of TL and FL segmentation. In a comprehensive review on AD~\cite{pepe2020}, Pepe et al. present and discuss different automatic and semi-automatic medical image analysis techniques for dissected aortic segmentation.
\section{Method}
\begin{figure} \centering
\includegraphics[width=\textwidth]{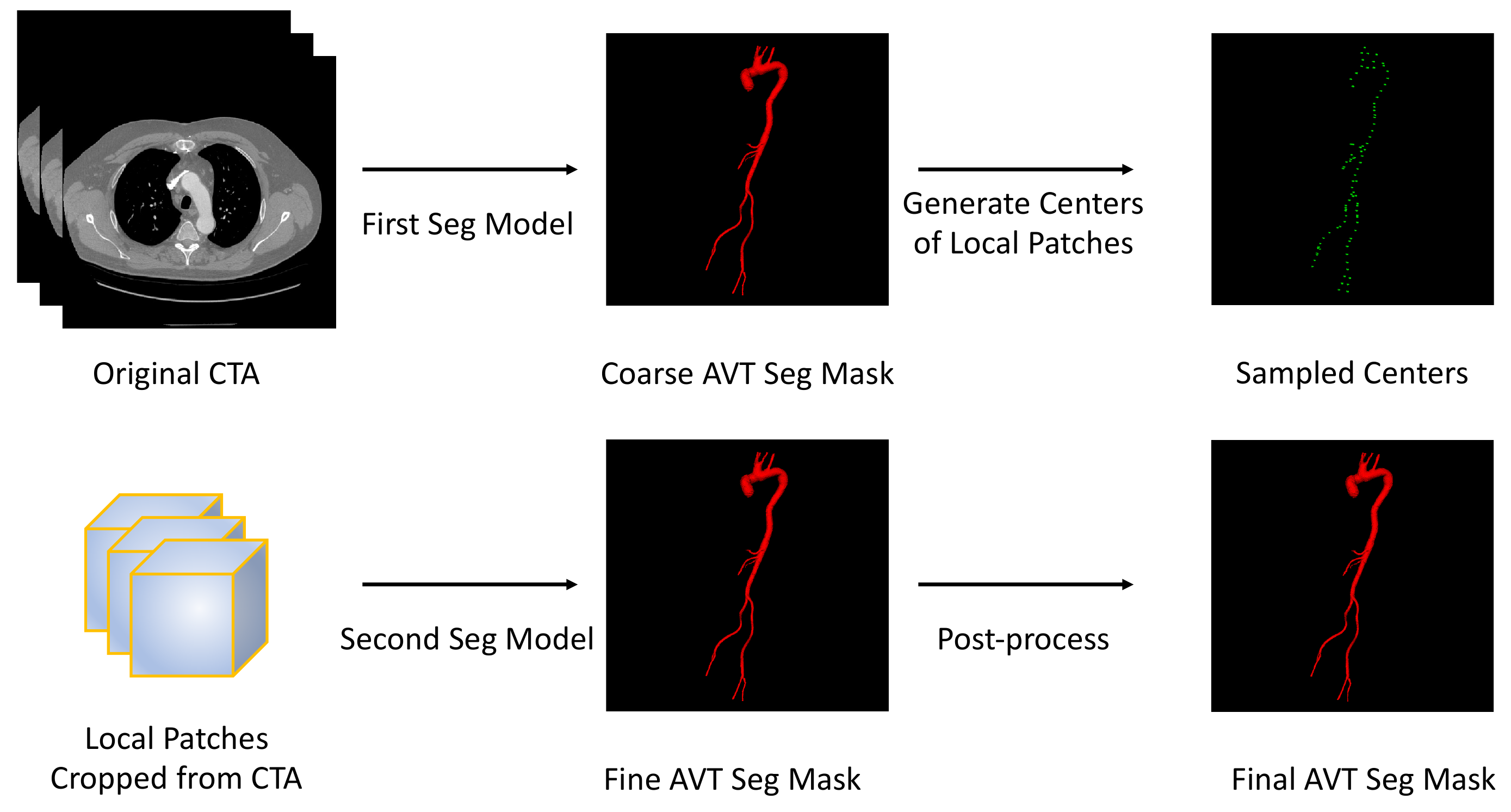}
\caption{The overview of the proposed \textbf{AVTSeg} method.} \label{fig1}
\end{figure}

The overview of the proposed \textbf{AVTSeg} method is shown in Fig.~\ref{fig1}.
\textbf{AVTSeg} consists of two segmentors, namely, the first and second segmentation models. Different from coarse-to-fine strategy (the input of the second model is cropped image from the output of the first model), two segmentors here are trained with the same spacing but with different patch size. The first segmentation model uses larger patch size and inference-stride for rapidly producing coarse AVT segmentation mask. Like region proposal networks in Faster R-CNN, the coarse AVT segmentation mask is also a type of attention. Then we generate a set of sparse centers of local patches along AVT. Note that a set of dense centers wastes computation because of much overlap between cubes. Additionally, centerline extraction of AVT using skeletonization or fast marching method is computation-intensive and time-consuming. Therefore, we use pseudo-centerline to sample positions of local patches. Along directions of transverse, sagittal and coronal plane, we find centroids of two-dimensional components and then merge into three-dimensional coordinates. The second segmentation model uses smaller patch size for attentively and accurately refining AVT segmentation mask.
\subsection{First Segmentation Model}
Fast segmentation of AVT by the fitst model is in order to accelerate medical image analysis pipeline and produce vesselness proposals. We adopt nnU-Net~\cite{isensee2021} as an incarnation of the first segmentation model because of its simplicity and powerfulness.
\subsection{Second Segmentation Model}
Accurate segmentation of AVT by the second model along the pseudo-centerline of AVT is in order to extract local vessel segment as complete as possible. In this stage, we also adopt nnU-Net~\cite{isensee2021} as the above first segmentation model but with different patch size and training patch sampling strategy.

\section{Experiments and Results}
\subsubsection{Datasets}
We tested \textbf{AVTSeg} method on 56 CTA scans from SEG.A. challenge 2023. This \textbf{SEG.A.} dataset consists of three centers, including the KiTS Grand Challenge (K dataset with 20 cases), the Rider Lung CT dataset (R dataset with 18 cases), and the Dongyang Hospital (D dataset with 18 cases)~\cite{radl2022}. We find that central intensities of clinically relevant targets in K dataset and R dataset are around at 1000, however, in D dataset, central intensity is 0. This inconsistent may hinder the training of neural networks. Therefore, we shift central intensities in K dataset and R dataset by subtracting 1024. The new version of \textbf{SEG.A.} dataset is \textbf{NEW.SEG.A}.

\subsubsection{Results}
As shown in Tab.~\ref{tab1}, we find that (1) the original dataset \textbf{SEG.A.} has the problem of inconsistent central intensity; (2) On both \textbf{SEG.A.} and \textbf{NEW.SEG.A} datasets, the proposed segmentation method in this experimental report get a promising segmentation performance in term of Dice and Jaccard similarity coefficients; (3) There are several corner cases in the competitive dataset (e.g., child's aorta in train-val data split 4) and this dataset has less cases for developing an accurate, generalizable and robust segmentation model.

\begin{table}[htbp]
\caption{Quantitative evaluation for AVT segmentation on original \textbf{SEG.A.} (A) and modified \textbf{NEW.SEG.A.} (B) datasets. DSC is Dice Similarity Score, HD is Hausdorff Distance, SS is Sobol' sensitivity, IoU is interaction over union.}
\begin{center}
\begin{tabular}{c |c |c c c c c c}
\hline
 & & \multicolumn{5}{c}{Mertics}\\
Method & dataset & DSC & HD & SS & IoU & Recall & Precision\\
\hline
\multirow{2}{*}{nnU-Net on fold0} & A & 86.8 & - & - & 80.6 & 86.6 & 90.8\\
 & B & \textbf{93.4} & - & - & \textbf{87.8} & \textbf{93.2} & \textbf{93.9}\\
\hline
\multirow{2}{*}{nnU-Net on fold1} & A & 92.8 & - & - & 86.7 & 94.3 & 91.7\\
 & B & \textbf{93.2} & - & - & \textbf{87.4} & 94.0 & \textbf{92.7}\\
\hline
\multirow{2}{*}{nnU-Net on fold2} & A & 94.2 & - & - & 89.1 & 93.2 & \textbf{95.4}\\
 & B & \textbf{94.8} & - & - & \textbf{90.2} & \textbf{94.5} & 95.3\\
\hline
\multirow{2}{*}{nnU-Net on fold3} & A & \textbf{95.0} & - & - & \textbf{90.6} & 95.0 & \textbf{95.2}\\
 & B & \textbf{95.0} & - & - & 90.4 & \textbf{95.3} & 94.8\\
\hline
\multirow{2}{*}{nnU-Net on fold4} & A & \textbf{84.5} & - & - & \textbf{75.9} & 82.8 & \textbf{91.0}\\
 & B & 83.7 & - & - & 74.8 & \textbf{84.0} & 86.4\\
\hline
\multirow{2}{*}{average performance} & A & 90.7 & - & - & 84.6 & 90.4 & \textbf{92.8}\\
 & B & \textbf{92.0} & - & - & \textbf{86.1} & \textbf{92.2} & 92.6\\
\hline
\end{tabular}
\label{tab1}
\end{center}
\end{table}

%
%
%
%

\end{document}